\documentclass[12pt]{iopart}
\usepackage{iopams}
\usepackage{graphicx}

\begin{document}

\title[QPS phenomenon in superconducting nanowires]{Quantum phase slip phenomenon in superconducting nanowires with low-Ohmic environment}
\author{J. S., Lehtinen$^{1}$ and K. Yu., Arutyunov$^{1,2}$}

\begin{abstract}
In a number of recent experiments it has been demonstrated that in
ultra-narrow superconducting channels quantum fluctuations of the order
parameter, alternatively called quantum phase slips, are
responsible for the finite resistance well below the critical temperature. The
acceptable agreement between those experiments and the models describing
quantum fluctuations in quasi-one-dimensional superconductors has been
established. However the very concept of the phase slip is justified when
these fluctuations are the relatively rare events, meaning that the effective
resistance of the system should be much smaller than the normal state
equivalent. In this paper we study the limit of the strong quantum
fluctuations where the existing models are not applicable. In particular case
of ultra-thin titanium nanowires it is demonstrated that below the expected
critical temperature the resistance does
not demonstrate any trend towards the conventional for a superconductor
zero-resistivity state even at negligibly small measuring currents.
Application of a small magnetic field leads to an unusual negative
magnetoresistance, which becomes more pronounced at lower temperatures. The
origin of the negative magnetoresistance effect is not clear.

\end{abstract}

\address{1 University of Jyvaskyla, Department of Physics, PB 35, 40014 Jyvaskyla, Finland}
\address{2 Nuclear Physics Institute, Moscow State University, 119992 Moscow, Russia}

\ead{janne.s.lehtinen@jyu.fi}

\maketitle

\section{INTRODUCTION}

Since the early years of superconductivity studies it has been noticed that
any superconducting transition $R(T)$ always has a finite width. Very often
sample inhomogeneity is the dominating factor. However with refinement of
fabrication methodology it became clear that even in the most homogeneous
samples the $R(T)$ transition is not infinitely narrow. The broadening of the
$R(T)$ dependence is most pronounced in systems with reduced dimensions where
the thermodynamic fluctuations have a larger effect. In particular case of
quasi-one-dimensional (1D) channels fluctuation-driven phase slips, the
concept first introduced by W. Little in 1967 \cite{PhysRev.156.396}, are
responsible for the finite resistivity in a narrow region below the critical
temperature \cite{PhysRev.164.498, PhysRev.172.427} $R(T)\sim exp(-F_{0}%
/k_{B}T)$ , where $F_{0}$ is the condensation energy of the smallest
statistically independent volume of the wire. Soon after formulation of the
model \cite{PhysRev.164.498, PhysRev.172.427} experiments
\cite{PhysRevLett.25.1180, PhysRevB.5.864} confirmed the validity of the
concept of the thermally activated phase slips (TAPS).

At lower temperatures the number of thermally activated phase slips is
exponentially suppressed and no resistance should be expected well below the
critical temperature. However later experiments
\cite{PhysRevLett.61.2137,Nature.404.971,PhysRevLett.97.017001,NanoLett.5.1029,PhysRevB.77.054508,PhysRevB.85.094508}
in extremely narrow superconducting nanowires have demonstrated the finite
resistance even at temperatures $T\ll T_{c}$. The effect has been associated
with quantum fluctuations of the order parameter, alternatively called
\textit{quantum phase slips} (QPS) \cite{PhysRevLett.78.1552, PhysRep.464.1}.
Yet another confirmation of the QPS reality came from the experiments studying
the persistent current in nanorings \cite{SciRep.2.213} resulting in building
a quantum two level system - qubit \cite{Nature.484.355}.

Though the quantum fluctuation phenomenon has already received the
experimental confirmation from several independent sources, the physics behind
is still under debates. Of special interest are the recent theoretical
developments predicting that a QPS junction is dual to a Josephson junction
\cite{NaturePhys.2.169, PhysRevLett.106.077004, PhysRevB.83.174511}. Contrary
to conventional transport measurements with low-Ohmic contacts
\cite{PhysRevLett.61.2137,Nature.404.971,PhysRevLett.97.017001,NanoLett.5.1029,PhysRevB.77.054508,PhysRevB.85.094508}%
, if a nanowire governed by QPSs ('QPS junction') is truly current biased,
then one should expect the development of the insulating state - the Coulomb
blockade \cite{PhysRevLett.108.097001} in a full accordance with a Josephson
junction in the regime of Bloch oscillations \cite{Zh.Eksp.Teor.Fiz.88.692,
PhysRevLett.67.2890, NaturePhys.2.169}

The mandatory pre-requisites of such a non-trivial experimental configuration
is the high enough rate of QPSs being of the same order as the experimentally
observed Coulomb gap. Contrary to such a sample current biased through
high-Ohmic electrodes, similar superconducting nanowire of the same diameter,
but with low-Ohmic probes (e.g. superconducting), should not demonstrate the
Coulomb gap, but its $R(T)$ and $I-V$ dependencies should show no traces of
'conventional' superconductivity either. The objective of this paper is to
study exactly the limit of extremely narrow nanowires where the QPS rate is
high enough to significantly affect the transport properties.

\section{THEORY BACKGROUND}

The impact of the quantum fluctuations on the shape of the superconducting
transition is qualitatively described by expression, similar to the TAPS
expression, \cite{PhysRep.464.1} $R(T)\sim exp(-F_{0}/h\Gamma_{QPS})$, where
instead of the thermal energy $k_{B}T$ stands the rate of QPS%

\begin{equation}
\Gamma_{QPS}\simeq\frac{R_{Q}}{R_{N}}\frac{\Delta(T)}{h}\left(  \frac{L}%
{\xi(T)}\right)  ^{2}\exp(-S_{QPS}), \label{Eq1}%
\end{equation}

where $R_{N}$ is the normal state resistance, $R_{Q}=h/2e= 6.45 k\Omega$ is
the superconducting quantum resistance, $L$ is the length of the wire,
$\Delta(T)$ and $\xi(T)$ are the temperature-dependent superconducting energy
gap and coherence length. $S_{QPS}=A({R_{Q}/R_{N})(L}/{\xi(T))}$ is the QPS
action, where constant $A$ is of the order of unit and cannot be more
precisely defined from the theory \cite{PhysRevLett.78.1552}. It is the only
true fitting parameter of the model, other parameters can be derived from the
experimental data. For our dirty limit \ samples $\xi\ll l$ the
superconducting coherence length can be estimated as $\xi\simeq\sqrt{l\xi
_{0}(T)}$, where $\xi_{0}\simeq\hbar v_{F}/\Delta(T)$ is the coherence length
in the clean limit $\xi_{0}\gg l$ and $v_{F}$ is the Fermi velocity. Utilizing
the text-book expression for the normal state resistance of a wire with
cross-section $\sigma$, length $L$ and resistivity $\rho_{N}$, one comes to a
conclusion that the effective resistance of a QPS-governed system
exponentially depends on the sample parameters $R(T)\sim exp(-\sigma
\sqrt{T_{c}}/\rho_{N})$. Hence thin nanowires of low-$T_{c}$ superconductors
with high resistivity (in normal state) are of advantage for observation of a
pronounced contribution of the QPS effect.

It should be noted that the model \cite{PhysRevLett.78.1552} is based on the
assumption that quantum fluctuations are relatively rare events. Or in other
terms, the corresponding QPS-related effective resistance of the nanowire
should be much smaller than the normal state resistance $R(T\ll T_{c})\ll
R_{N}$. As it comes from the formulated above objective of the paper, the
limit of strong fluctuations violates this requirement. Unfortunately, to our
best knowledge, the limit of strong quantum fluctuations has not yet been
properly treated theoretically. Hence, the expression (\ref{Eq1}) can be
considered only as a certain guideline. As it will be shown below, indeed when
the contribution of the QPS effect on the shape of the $R(T)$\ transition is
already noticeable, the further reduction of the nanowire cross-section
$\sigma$ leads to complete flattening of the $R(T)$ dependence not described
by the model \cite{PhysRevLett.78.1552}.

\section{RESULTS and DISCUSSION}

The research related to mesoscopic superconducting is limited by the
fabrication technology. In particular case of QPS effect, to reach the regime
of interest $h\Gamma_{QPS}\lesssim\Delta$ the nanowire should have very small
cross-section. For example for aluminum the cross-section below 10 nm (still
maintaining the high level of uniformity!) is mandatory \cite{NanoLett.5.1029}%
. Following Eq. (1), for materials like Nb, Sn or Pb with relatively high
critical temperature one gets even more pessimistic estimations approaching 1
nm scale. This note explains why in particular case of niobium no traces of
the QPS effect have been observed down to 7 nm scales \cite{APL.83.512}. With
proper material selection the extremely tough fabrication requirements can be
somehow relaxed. From our previous studies we have already learnt that in
titanium the QPS effect is observable at sub-40 nm scales
\cite{PhysRevB.85.094508,SciRep.2.213}, which is achievable with the standard
e-beam lithography technique. In addition, titanium is an easy to work
material and the extended microscopic and elemental analysis reveals no severe
structural defects or/and impurities \cite{PhysRevB.85.094508}. For the
purpose of the present work we prepared several titanium nanowires with length
$L=10~\mu m$ and the effective diameter $\sqrt{\sigma}$ between 27 and 48 nm.
The structures were fabricated with the standard lift-off e-beam lithography,
and the titanium was deposited in UHV e-beam evaporator at residual pressure
$\sim5\ast10^{-9}$ mbar. The substrate covered with exposed PMMA/MAA mask was
cleaned with low-energy $O_{2}$ plasma immediately before the metal
evaporation. Based on the measured resistivity $\rho_{N}=0.5\ast10^{-6}$
$\Omega\cdot m$ , which is comparable with the tabulated value for the clean
bulk titanium, the quality of the metal thin film is acceptable. Previous TEM
and TOF-ERDA analyses of titanium nanostructures fabricated using identical
conditions and parameters confirmed the high material quality: the highest
concentration of foreign inclusions inside the titanium matrix corresponds to
$\sim0.4~at$ $\%.$ of oxygen \cite{PhysRevB.85.094508}. This low concentration
of impurities, combined with TEM analysis, disables any speculations about the
presence of sample inhomogeneities capable to significantly broaden the $R(T)$
dependence. The SPM analysis of the studied samples (Figure 1) confirmed the
'standard' sample quality with the surface roughness of about several nm
\cite{PhysRevB.85.094508, SciRep.2.213,APA.79.1769,Nanotechnology.19.055301}.
The mean free path $l$ \ for the wire can be evaluated from the measured
normal state resistance $R_{N}$\ and the known material constant $l\rho
_{N}\approx10\ast10^{-16}~\Omega m^{2}$ varying slightly in different
literature sources. For our samples estimations give the mean free path
$l\simeq$ 1 to 2 nm. The result agrees well with the SPM and TEM measured grain
size of roughly 2 to 3 nm.

\begin{figure}[ptb]
\includegraphics[width=80mm]{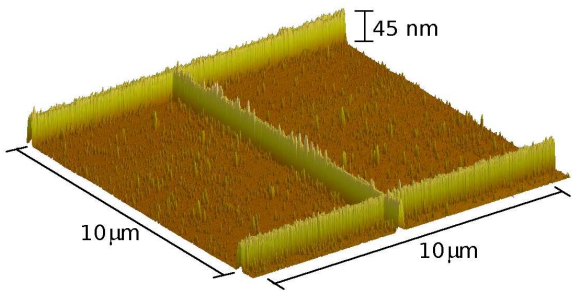}\caption{SPM image of a typical
all-titanium nanostructure on Si/Si$O_{x}$ substrate.}%
\end{figure}

The $R(T)$ measurements (Figure 2) were made employing the conventional
four-probe configuration using 7 Hz lock-in technique at rather small
excitation currents of about 50 pA. The experiments were made in $He^{3}%
He^{4}$ dilution refrigerator with the base temperature down to 17 mK. All
input/output lines were carefully protected from the noisy electromagnetic
environment utilizing multi-stage RLC filtering enabling the electron
temperature $\sim$35 mK at a base (phonon) temperature $\sim$20 mK
\cite{PhysRevB.83.104509}. For sufficiently thick nanowires with the effective
diameter $\sqrt{\sigma}\gtrsim$ 40 nm the shape of the $R(T)$ dependencies
follows our earlier findings: pronounced superconducting transition with\ the
low temperature data $R(T)<R_{N}/10$ which can be described by the QPS model
\cite{PhysRevLett.78.1552} with the realistic set of fitting parameters
\cite{PhysRevB.85.094508}.\ However the thinner the sample, the less the
resistance $R(T)$ drops below the certain 'critical temperature' (Figure 2).
It should be noted that for titanium the critical temperature $T_{c}$
decreases with the decrease of the cross-section $\sigma$. The size dependence
of the critical temperature for low dimensional superconductors is a
well-known effect, though the commonly accepted explanation of the phenomenon
still does not exist \cite{PhysRevB.74.052502, PhysRevLett.83.191}. Thus the
accurate experimental definition of the critical temperature for the thinnest
samples $T_{C}(\sigma)$, necessary for the theory fitting (Equation
\ref{Eq1}), is rather problematic due to the strongly broadened $R(T)$
transition. Nevertheless for the thinnest samples $\sqrt{\sigma}\lesssim$ 35
nm no set of realistic QPS model parameters can fit the experimental data
(Figure 2). For these ultra-thin samples the experimental $R(T)$ dependency is
so weak, that the model \cite{PhysRevLett.78.1552} applicability criterion
$R(T)\ll R_{N}$ is not satisfied down to the lowest experimentally obtainable temperatures.

\begin{figure}[ptb]
\includegraphics[width=80mm]{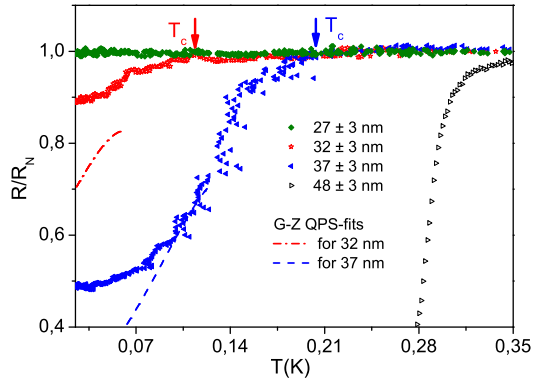}\caption{Temperature dependence of
the resistance of the four titanium nanowires with length $L=10~\mu m$ and the
effective diameters indicated in the inset. Model \cite{PhysRevLett.78.1552}
calculations using the realistic fitting parameters $l=$ 1 nm, $v_{F}%
$=1.79$\times10^{6}$ m/s, $A$=0.098 and the critical temperature $T_{c}$=200
mK and 115 mK for the 37 nm and 32 nm samples, respectively, are plotted with
dashed lines. Best fit $T_{c}$'s are indicated with arrows. Clearly the model
\cite{PhysRevLett.78.1552}, applicable in the limit $R(T)\ll R_{N}$,\ fails to
provide any acceptable fits for the thinnest samples $R(T)$ data.}%
\end{figure}

At temperatures well-below the critical temperature $T\ll T_{c}$ the
$dV/dI(I)$ dependencies (Figure 3) for the thickest studied samples
demonstrate the conventional destruction of the zero-resistance state by
current with the well-pronounced critical current $I_{c}$. The double shape of
the transition (e.g. at $\sim$22 nA and $\sim$28 nA for the 48 nm sample)
presumably originates from slightly different effective cross-section of the
samples at the node regions. With decrease of the nanowire cross-section, as
expected from the $R(T)$ dependencies, the the zero-resistance state
disappears and only some residual 'critical current' peculiarities can be
traced. In the thinnest samples the zero-biased differential resistance
$dV/dI(I\rightarrow0)$\ very slightly increases indicating the presence of a
weak Coulomb blockade. Observation of a pronounced Coulomb effects requires
the true current biasing (including the high frequencies!)
\cite{PhysRevLett.108.097001} and this experimental realization is not within
the scope of the present paper.

\begin{figure}[ptb]
\includegraphics[width=80mm]{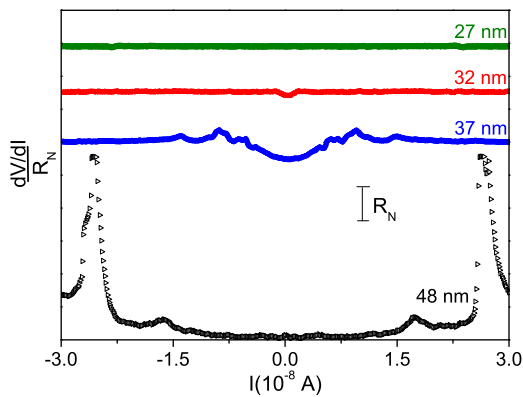}\caption{Differential resistance
dV/dI, normalized by the normal state resistance $R_{N}$, as function of the
bias current I for the same four samples as in Figure 2. The data for the 48
nm nanowire (open right-pointing triangles) have been collected at 80 mK,
while for the other three samples - at 20 mK. For clarity the curves are
vertically shifted. The scale corresponding the normal state resistance
$R_{N}$ is indicated with vertical bar.}%
\end{figure}

Yet one more interesting observation is the negative magnetoresistance (nMR)
which is observed in the thinnest samples with pronounced QPS contribution
(Figure 4). The effect increases with lowering the temperature. Similar
phenomenon has been reported in aluminum \cite{PhysRevB.77.054508}, niobium
and $MoGe$ \cite{PhysRevLett.97.137001} and lead \cite{PhysRevLett.78.927}
nanowires. The origin of the nMR in these quasi-1D channels is still under
debates. In case of niobium and $MoGe$ it was conjected that some rogue
magnetic moments might be present, and their pair breaking contribution,
active at lower magnetic fields, is suppressed by higher fields leading to the
observed nMR \cite{PhysRevLett.97.137001}. However this mechanism needs an
independent proof of the existence of these magnetic moments, and cannot
explain the nMR in such material as aluminum \cite{PhysRevB.77.054508}, where
it is a well-known fact that majority of magnetic impurities can obtain a
non-negligible magnetic moment only at relatively high concentrations
\cite{JAP.98.016105}. Another explanation of the nMR phenomenon deals with the
magnetic field suppression of the charge imbalance accompanying each phase
slip event \cite{PhysicaC.468.272}. Somehow related mechanism, capable to
provide nMR, deals with more effective suppression of superconductivity in
(wider) electrodes affecting the phase slip formation in the (thinner)
nanowire \cite{PhysRevB.75.184517}.

\begin{figure}[ptb]
\includegraphics[width=80mm]{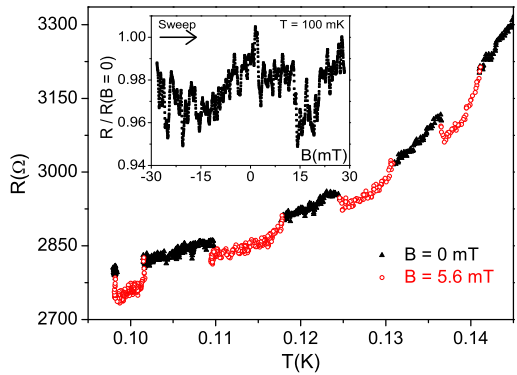}\caption{Temperature dependence of
the resistance of the titanium nanowire with length L = 30 um and the
effective diameter $39 \pm3 ~ nm$ measured at zero magnetic field (filled
triangles) and in perpendicular magnetic field B = 5.6 mT (open circles). Both
data were obtained while the same slow temperature sweep just by switching on
and off the magnetic filed. The inset shows the representative region with the
negative magnetoresistance, measured at a constant temperature T = 100 mK. }%
\end{figure}

\section{CONCLUSIONS}

$R(T,B)$ and $V-I$ characteristics of titanium nanowires with sub-50 nm
diameters were measured. The shape of the $R(T)$ dependencies for the
'not-too-narrow' samples with the effective diameters $\sqrt{\sigma}\gtrsim$
40 nm confirms the earlier findings: pronounced superconducting transition
with\ the low temperature data $R(T)<R_{N}/10$ which can be described by the
quantum phase slip model \cite{PhysRevLett.78.1552}. In thinner samples the
$R(T)$ transitions dramatically flatten disabling any comparison with the
existing fluctuation models, which assume that the phase slips are still the
rare events and hence the effective resistance should be much smaller than the
normal state resistance. The $dV/dI(I)$ dependencies confirm the $R(T)$ data
conclusion about the absence of the truly zero-resistance state in the
thinnest samples. The negative magnetoresistance is observed, while the origin
of the effect is not clear.

\ack
The National Graduate School in Materials Physics (NGSMP, Finland) is
gratefully acknowledged for financial support.

\end{document}